\documentclass{sigchi-ext}
\usepackage[T1]{fontenc}
\usepackage{textcomp}
\usepackage[scaled=.92]{helvet} 
\usepackage{graphicx} 
\usepackage{balance}  
\usepackage{booktabs} 
\usepackage{ccicons}  
\usepackage{ragged2e} 
\usepackage{subfigure}
\usepackage{tabularx}
\usepackage{amsmath}
\usepackage{fixltx2e}
\usepackage[noadjust]{cite}
\graphicspath{{images/}}
\usepackage{url}
\usepackage{multirow}
\usepackage{color}
\usepackage{refcount}
\usepackage{epstopdf}

\def\plaintitle{SIGCHI Extended Abstracts Sample File: Note Initial
  Caps} 
\def\emptyauthor{}
\def\plainkeywords{Authors' choice; of terms; separated; by
  semicolons; include commas, within terms only; required.}

\title{Why Did They \#Unfollow Me?\\ Early Detection of Follower \\ Loss on Twitter}

\numberofauthors{3}
\author{%
  \alignauthor{\textbf{Suman Kalyan Maity}\\
       \affaddr{Dept. of CSE}\\
       \affaddr{IIT Kharagpur, India}
       \email{sumankalyan.maity@cse.iitkgp.ernet.in}}
       \vfil \alignauthor{\textbf{Ramanth Gajula}\\
       \affaddr{Dept. of CSE}\\
       \affaddr{IIT Kharagpur, India}
       \email{ramanth139@gmail.com}}
       \vfil \alignauthor{\textbf{Animesh Mukherjee}\\
       \affaddr{Dept. of CSE}\\
       \affaddr{IIT Kharagpur, India}
       \email{animeshm@cse.iitkgp.ernet.in}}}

\definecolor{linkColor}{RGB}{6,125,233}
\hypersetup{%
  pdftitle={\plaintitle},
  pdfauthor={\emptyauthor},
  pdfkeywords={\plainkeywords},
  bookmarksnumbered,
  pdfstartview={FitH},
  colorlinks,
  citecolor=black,
  filecolor=black,
  linkcolor=black,
  urlcolor=linkColor,
  breaklinks=true,
}

\begin{document}

\CopyrightYear{2018} 
\setcopyright{rightsretained} 
\conferenceinfo{GROUP '18}{January 7--10, 2018, Sanibel Island, FL, USA}\isbn{978-1-4503-5562-9/18/01}
\doi{https://doi.org/10.1145/3148330.3154514}
\copyrightinfo{\acmcopyright}

\maketitle

\RaggedRight{}

\begin{abstract}
Having more followers has become a norm in recent social media and micro-blogging communities. This battle has been taking shape from the early days of Twitter. Despite this strong competition for followers, many Twitter users are continuously losing their followers. This work addresses the problem of identifying the reasons behind the drop of followers of users in Twitter. As a first step, we extract various features by analyzing the \textit{content of the posts} made by the Twitter users who lose followers consistently. We then leverage these features to early detect follower loss. We propose various models and yield an overall accuracy of 73\% with high precision and recall. Our model outperforms baseline model by 19.67\% (w.r.t accuracy), 33.8\% (w.r.t precision) and 14.3\% (w.r.t recall).
\end{abstract}

%
\category{J.4}{Computer Applications}[Social and Behavioral Sciences]
\category{K.4.2}{Computers And Society}[Social Issues]

\section{Introduction}
Followership of users in social media is an important factor since it indicates social prestige and popularity for the users. Followers have a proportional impact on how far and wide one's message spreads and the rate at which one can get social recognition in form of reposts, shares, likes etc.\footnote{https://blog.bufferapp.com/definitive-guide-social-media-metrics-stats} It helps in outreach, helps in forming new social relationships. Though people have studied followership gain, there are very few studies that looked into the other side of the spectrum of this online relationship - the ``unfollowing'' behavior. Like gain in followership, followership loss has also important social connotation and business implications. 
\marginpar{
 \fbox{
   \begin{minipage}{0.93\marginparwidth}
\textbf{Dataset preparation}
We construct our dataset through web-based crawls of the profile information of 9.3 million users at two different time points -- i) June 2014 and ii) September 2016. We then select those users who have at least 1000 followers in June 2014 and lost some followers by September 2016. We create two datasets based on followership gain/loss characteristics. \textit{Dataset1} consists of users who lost at least 30\% of their followers. \textit{Dataset2} has users who lost at most 2\% followers plus users who gained at most 2\% followers. Our objective here is to identify features that discriminate this set (which corresponds to mostly accidental loss/gain of followers) from the set of users who incur a real loss of followers (i.e., \textit{dataset1}). We further remove those users who did not tweet in English. We then randomly sample out 8000 users from \textit{dataset1} and a similar number of users from \textit{dataset2} for the subsequent study.
\end{minipage}}\label{dataset}}
Twitter or other social media are extensively used by media houses, various industry outlets from technology to fashion, political personalities. Therefore, followership loss of such entities could mean decrease in face value and which could directly/indirectly impact business. For instance, boxer cum politician, Manny Pacquiao lost 2 million followers over his gay comments\footnote{http://bit.ly/2yFnnHF}, Indian prime minister Narendra Modi reportedly lost 313,312 followers after announcing demonetization of 500 and 1000 notes\footnote{http://bit.ly/2gxAzIs}. In our dataset containing 9.3 million Twitter users, 26\% of the users are found to have suffered a net loss in a two years span. For instance, a user from our dataset who had 114K followers, tweeted only about mundane details of his day to day activities and therefore lost 85 percent of his followers. Another user who had 176K followers, lost 55\% of his followers most likely because the tweets mostly portray political propaganda and the tweet frequency is as high as $\sim 200$ tweets per day. Though both gain/loss in followership can be contextual to different relationships and situations, however, in this work we try to find holistically what factors - like \textit{social behavior}, \textit{textual content of posts}, \textit{language usage and network structure} - lead to follower loss.

\indent \textbf{Related work}: There have been few studies done by researchers to understand the dynamics of unfollowing in various OSNs. Kwak et al.~\cite{ICWSM124598} reported that 43\% of active users unfollow at least once during 51 days. Twitter users have unfollowed those users who left many tweets within a short time, created tweets about uninteresting topics, or tweeted about the mundane details of their lives~\cite{Kwak:2011:FOR:1978942.1979104,Moon2011}. Also Twitter users appreciate receiving more attention than giving when it comes to mentions, retweets etc., and this is pronounced in the act of unfollow~\cite{ICWSM124598}. Another popular mode of unfollowing in Twitter is burst unfollowing~\cite{DBLP:journals/corr/MyersL14}. In this work, we propose models for early prediction of loss of followers on Twitter mainly focusing on the content and the language usage in the tweets posted by the users. In specific, we make use of the activities and the content of the tweets of the victim (i.e., the person losing the followers) only. Building such a model would enable the victim early in time to know the specific online behavior that could result in followership loss. Having such succinct clues can guide the victim as to how to contain his/her behavior to avoid loss of followers (which is usually very hard to accumulate).  
\section{Factors behind follower loss}\label{factors}
The factors below attempt to extract the textual content and the language usage behavior of the victims.

\noindent\textbf{Use of offensive/profane words in tweets}: We use a list of offensive and profane words from \url{https://www.cs.cmu.edu/~biglou/resources/bad-words.txt} and manually label their offensive/badness score. We calculate badness influence per tweet as the sum of badness of the words used in the tweet normalized by the number of words in the tweet. The average badness influence of all the tweets of a user gives the \textit{badness coefficient} of a user.

\noindent\textbf{Repetitive content: word diversity}: Repetitive content is considered generally as boring in social media, unless the content is very trendy. Let \textit{T\textsubscript{\textit{u}}} be the multi-set of words from all the filtered tweets of user \textit{u} and \textit{W} be the set of all unique words in \textit{T\textsubscript{\textit{u}}} and \textit{p(w$|$T\textsubscript{\textit{u}})} be the probability of word 
\textit{w} belonging to T\textsubscript{\textit{u}}. We now define content diversity as $ContentDiv(u)=- \sum_{w\in W} p(w|T\textsubscript{\textit{u}})\times log(p(w|T\textsubscript{\textit{u}}))$

\noindent\textbf{Topic diversity}: Topic diversity also captures the notion of repetitive content by finding topics in the tweets of user rather than directly using the words. We use LDA~\cite{Blei:2003:LDA:944919.944937}, for the discovery of latent subtopics and calculate the topical diversity for an user $u$ as 
$TopDiv(\textit{u})=- \sum_{\textit{k}=1}^{\textit{K}} p(topic\textsubscript{\textit{k}}|T\textsubscript{\textit{u}})\times log(p(topic\textsubscript{\textit{k}}|T\textsubscript{\textit{u}}))$ where $T_u$ denotes the set of tweets as a document for user $u$.


\marginpar{
 \fbox{
   \begin{minipage}{0.98\marginparwidth}
\textbf{Tweet bursts}: Bursts of tweets sometimes draw an unwanted attention in Twitter. An example of tweet burst includes a long story posted as a continuum of tweets. Twenty out of 22 respondents reported that they unfollowed 39 people because of burst tweets~\cite{Kwak:2011:FOR:1978942.1979104}. Consider the array \textit{T\textsubscript{\textit{u}}} of tweets of a user \textit{u} sorted according to the tweet arrival time. We define tweet burst as a maximal sub-array $ T\textsubscript{\textit{u}}[i..j] \mid \forall k , i \leq k < j , t(k+1)-t(k) \leq 1000
$ where $t(k)$ denotes the arrival time of the \textit{k}$^{th}$ tweet of the user \textit{u}. Time period of a burst $T\textsubscript{\textit{u}}[i..j]$ is defined as $t(j)-t(i)$. In the equation 1000 is a chosen hyper parameter. We use the following set of features extracted from tweet bursts - \textit{mean inter-burst arrival time, avg. time period of a burst, max. time period of a burst, minimum time period of a burst, no. of bursts}.\\
\noindent\textbf{Tweet length}: We observe that users with tweets having very short length and users with tweets occupying most of the allowed space, are more likely to lose followers.\\
\end{minipage}}\label{tweetburst}
 }


\noindent\textbf{Tweet rate}: In Twitter, users would hardly want their feeds to be overflown by the tweets from a single other user. We capture this notion using the rate at which a given user is tweeting which simply is the time difference of the first and the last tweets of the user normalized by the total number of tweets so far (in the data) of the user.

\noindent\textbf{Mentions per tweet}: We calculate \textit{MentionCoeff} as the average number of mentions per tweet. Users who \textit{mention} infrequently are able to less engage other users and might get unfollowed.


\noindent\textbf{Mention entropy}: The \textit{MentionCoeff} measure might implicitly (and incorrectly) indicate that a particular user \textit{u}, who mentions only a small set of people very frequently, is very less probable of losing followers. However, these users might also be prone to losing followers. For example, users who follow 1000 people communicate with only about 70 people on average~\cite{Kwak:2011:FOR:1978942.1979104}. We capture this notion by using \textit{MentionEntropy}. Let \textit{M} be the list of distinct users mentioned by user \textit{u} and let \textit{p(m  $\vert$ u)} denote the probability with which user \textit{u} mentions user \textit{m} in his/her tweets.
So, $MentionEntropy(u)=- \sum_{m\in M} p(m|u)\times log(p(m|u))$

Users who have a low mention entropy are more from \textit{dataset1} indicating that users engaging only a particular set of other users in their tweets repeatedly are prone to lose more followers in future.


\noindent\textbf{Usage of urls in tweets}: Urls are popular in Twitter community for redirection. However, excessive usage of urls is usually not encouraged in the community because that is often interpreted as spamming. Users of \textit{dataset1} have more average url count per tweet indicating that people who use excessive urls are prone to loss of followers. 



\noindent\textbf{Profile description and verification status of user}: Profile description renders authenticity to a user profile. Interestingly, in our dataset, users who had profile description were less likely to lose followers. Verification status is also an important factor. Verified users usually have a net gain of followers. 90\% of the total verified users gained followers in our dataset. 

\noindent\textbf{Network features} We have constructed the following two networks - a) mention network b) content similarity network of the users in the dataset.\\
\textbf{a) Mention Network}: We consider the mention network of users in the full dataset where the nodes are the users and a directed edge is created from $a$ to $b$ if $a$ mentions $b$ at least once in his/her tweets. Only those users who have their (in-degree + out-degree) $>$ 0 are included in the network. $\sim$17\% of users from both datasets combined are present in this mention network. We have used various centrality and clustering based features (appropriately scaled) --\textit{in-degree centrality, out-degree centrality, betweenness centrality, closeness centrality, eigenvector centrality, clustering coefficient} from this network. 

\textbf{b) Content similarity network}: We consider the tweets of the users as bag-of-words. We then compute user-user similarity using the Jaccard co-efficient between the tweets. We then construct a network with nodes as users and edges indicating similarity between word usage of users. Through inspection of the distribution of similarities, we prune those edges with similarity values less than 0.3. In the resulting graph, the similarity feature for a user is extracted as follows: for a node $n$, all the neighboring nodes whose corresponding users are in the training set are considered and the majority class of neighbors is used as a feature. Clustering coefficient of similarity network is also used as a feature.

\marginpar{
 \fbox{
   \begin{minipage}{0.98\marginparwidth}
\noindent\textbf{Psycholinguistic aspects of tweets}\\
We also perform psycholinguistic analysis of the tweets to observe if there exists any pattern leading to follower loss. The cognitive, linguistic and psychological dimensions are captured through different categories provided by the LIWC tool~\cite{liwc}. There are 64 different categories that LIWC extracts from the tweet texts. First, we collect the words related to each of these 64 categories. Next, we find for each category \textit{c}, the number of words in the tweets of user $u$ which belong to the category \textit{c} and normalize this value by the total number of tweets of the user \textit{u}. Some of the key points to note here are that users who lose more followers use more negation words, less inclusive words as well as less insightful words.
\end{minipage}}\label{liwc}}
\section{Prediction Framework}
In this section, we present our model to early predict follower loss. Apart from the content based features, we have also used the following features - no. of followers, no. of followees, followee/follower ratio.\\
\vspace{1mm}
\noindent \textbf{Baseline Model:} In previous studies, many link based features from the follower-followee network like homophily, link exchange, follower overlap, tie strength \cite{ICWSM124598} have been used as factors for followership loss. We create a baseline model by only using those features which are from the perspective of user who get unfollowed and we shall compare our model with this baseline model.\\
\vspace{1mm}
\noindent\textbf{Our Model: Doc2vec + features} Apart from the features described above, we obtain vector representation of users using the state-of-art Doc2vec model\cite{DBLP:journals/corr/LeM14}. The word vectors are trained from the dataset of tweets. We feed the user vectors along with the features to feed-forward multi-layer perceptron (MLP) and train using cross-entropy loss for the classification task. We perform a 10-fold cross validation to evaluate our model. We vary the values of $K$\footnote{best result for $K = 30$} (number of topics in LDA) and other hyper parameters to obtain the best results.


\begin{table}[h]
\begin{center}
\vspace{-1mm}
\caption{Evaluation results.}
\vspace{-1mm}
\label{tab:results}
\resizebox{\columnwidth}{!}{%
\begin{tabular}{||c|c|c|c|c|c||}
	\hline
    Models  &Accuracy &Precision &Recall  &F1-score &ROC-area \\\hline
    Baseline Model &61\%&0.65 &0.70 &0.67 &0.62\\\hline 
    \textbf{Our Model} &\textbf{73\%} & \textbf{0.73} &\textbf{0.87} & \textbf{0.80} & \textbf{0.71}\\\hline
    %
\end{tabular}
}
\end{center}
\vspace{-2mm}
\end{table}

\textbf{Results:} Table~\ref{tab:results} summarizes the results. Our model significantly outperforms the baseline model by 19.67\% (w.r.t accuracy), 33.8\% (w.r.t precision) and 14.3\% (w.r.t recall). To understand which features are discriminative we rank them by their $\chi^2$ values. The top six discriminative features came out to be - \textit{avg. tweet burst time period, max. tweet burst time period, tweet frequency, mention entropy, topic diversity, eigenvector centrality of mention graph}.

\section{Conclusions and implications} 
In this paper, we identify various socio-linguistic factors behind followership loss and propose a feature-based model for followership loss prediction that achieves a good accuracy of 73\% and significantly outperforms the baseline model. The most discriminative factors are related to the users' tweeting behavior - frequency of tweets, their burstiness, the engaging ability of the user and the topic diversity of the user's tweets.

Our research can be helpful for Twitter users in various ways - i) to early identify the followership loss in near future ii) enabling victims to quickly take corrective measures/actions to stop the trend of follower loss and iii) help the Twitter service as a whole to build a ``tweet-properly''-like recommendation system for the subscribers to help them avoid unforeseen follower loss.  

\clearpage
\bibliographystyle{SIGCHI-Reference-Format}
\bibliography{ref}

\end{document}